\documentclass{PoS}

\usepackage{aas_macros}
\usepackage[numbers]{natbib}
\usepackage{amsmath}
\usepackage{amssymb}
\usepackage{txfonts}
\usepackage{setspace}
\usepackage{color}
\usepackage{graphicx}
\usepackage{fancyhdr}

\newcommand{\twofigureoutpdf}[3]{\centerline{}
   \centerline{\includegraphics[width=3.05truein]{#1}
        \hspace{-0.05truein} \includegraphics[width=3.05truein]{#2}}
        \vspace{-0.05truein}
    \caption{#3} }    

    \newcounter{mybibitem}
    \makeatletter
\def\@biblabel#1{\stepcounter{mybibitem}[\themybibitem]\,  }
   \makeatother
   
\title{High-Energy Emission from Black Widows and Redbacks}

\ShortTitle{MSP Binaries}

\author{\speaker{Zorawar Wadiasingh}\\
        Centre for Space Research, North-West University Potchefstroom\\
        E-mail: \email{zwadiasingh [at] gmail.com}}

\author{Alice K. Harding \\
NASA Goddard Space Flight Center \\
E-mail: \email{ahardingx [at] yahoo.com} }

\author{ Christo Venter \\
Centre for Space Research, North-West University Potchefstroom \\
E-mail: \email{christo.venter [at] nwu.ac.za} }

\author{ Markus Böttcher \thanks{NRF SARChI Chair} \\
Centre for Space Research, North-West University Potchefstroom\\
        E-mail: \email{Markus.Bottcher [at] nwu.ac.za }
}

\abstract{ A large number of new black widow and redback energetic millisecond pulsars with irradiated stellar companions have been discovered through radio searches of unidentified \emph{Fermi} sources.  We construct a 3D emission model of these systems to predict the high-energy emission
components from particles accelerated to several TeV in the intrabinary
shocks, and its predicted modulation at the binary orbital period. Synchrotron emission is expected at X-ray
energies and such modulated emission has already been detected by
\emph{Chandra} and \emph{XMM-Newton} in some systems. Synchrotron and inverse Compton emission from relativistic particles in the pulsar wind and intrabinary shock can probe the unknown physics of pulsar winds and relativistic shock acceleration in these compact binaries. Orbitally-modulated emission in the GeV and TeV bands may be detectable under some favorable conditions.}

\FullConference{3rd Annual Conference on High Energy Astrophysics in Southern Africa  -HEASA2015,\\
		18-20 June 2015\\
		University of Johannesburg, Auckland Park, South Africa}

\begin{document}

\vspace{-0mm}
\section{BACKGROUND}
\vspace{-3mm}

Recycled Galactic millisecond pulsars (MSPs) with low-mass companions are an old  $\gtrsim$ Gyr subset of the pulsar population which are relatively stable astrophysical laboratories for a variety of high-energy astrophysical phenomena and physics. Since the launch \emph{Fermi}, the number of these systems has burgeoned from a handful to almost 30\footnote{https://confluence.slac.stanford.edu/display/GLAMCOG/Public+List+of+LAT-Detected+Gamma-Ray+Pulsars}. These binaries are predominantly identified by precision radio timing of the MSP, revealing the existence of a companion in a circular orbit with period $<1$ day through a systematic radial velocity variation in the MSP timing solution, as was established with the first source in this class PSR B1957+20 \cite{1988Natur.333..237F}.  The MSP steadily irradiates and ablates the tidally-locked secondary star with an energetic $\sim 10^{34-35}$ erg s${}^{-1}$ pulsar wind, and a stable relativistic intrabinary shock is anticipated between the two stars \cite{1988Natur.333..832P}. An empirical classification based on the companion mass $M_c$ \cite{2011AIPC.1357..127R} roughly partitions the MSP binaries into black widows (BWs, $M_c \lesssim 0.05 M_\odot$) and redbacks (RBs, $M_c \gtrsim 0.1 M_\odot$) (colloquially ``spider'' or  ``Latrodectus'' binaries, since the recycled MSP state is evolved by angular momentum transfer from devouring the companion) that loosely traces the influence of the companion relative to the MSP. The RB companions are typically non-degenerate and resemble main sequence stars in their optical spectra, while many BW secondaries are more unusual.   A few systems have been found that vacillate between an accretion-powered low-mass X-ray binary state, where the pulsar is shrouded, and a MSP wind-powered RB state \cite{2009Sci...324.1411A, 2013Natur.501..517P}, providing strong evidence for the evolutionary link between these two binary classes.

The energetics of BWs and RBs are dominated by the pulsar wind, produced by the conversion of rotational energy and angular momentum of the MSP into an electromagnetic and particle wind. The pulsar wind is generally thought to be Poynting-flux dominated, with magnetization parameter $\sigma \gtrsim 1$, upstream of the intrabinary shock. At the relativistic intrabinary shock, a significant fraction of the wind energy is expected to be converted into particle energy and radiation  \cite{1990ApJ...358..561H, 1993ApJ...403..249A} possibly up to TeV energies for particles. There may also be relativistic particle acceleration and conversion to particle dominance $\sigma < 1 $ within pulsar winds \cite{2012Natur.482..507A}, a scenario which could be tested with nearby BWs and RBs. Since the orbital length scales of typical BWs and RBs are $\sim 10^{11}$ cm, these systems probe physical regions of a pulsar wind much smaller than the typical termination shock in young pulsar wind nebulae. 

Moreover, these BW and RB systems can possess multifaceted multiwavelength observational characteristics that can constrain several crucial orbital and pulsar parameters, thereby affording predictive power to physical models. For instance, optical modulations of the companion with a physically-motivated model of anisotropic heating can tightly constrain the system inclination and companion size \cite{2007MNRAS.379.1117R, 2013ApJ...769..108B} while frequency-dependent radio eclipses at superior conjunction of the MSP can constrain the extent of the shocked companion wind. If the pulsar and orbital rotation axes are aligned, modeling and fitting of the outer-magnetospheric pulsed $\gamma$-ray (and radio) light curves can yield an estimate of the angle between the pulsar spin axis and observer direction $\zeta$ and hence inclination $i$  \cite{2014ApJS..213....6J}. We schematically illustrate the geometry of a typical MSP binary in Figure 1. In this paper, we report on our initial findings from our 3D emission model, taking into account the system inclination and radio eclipses to constrain the intrabinary shock.

\begin{figure}[t]
\centering
\includegraphics[scale=0.4]{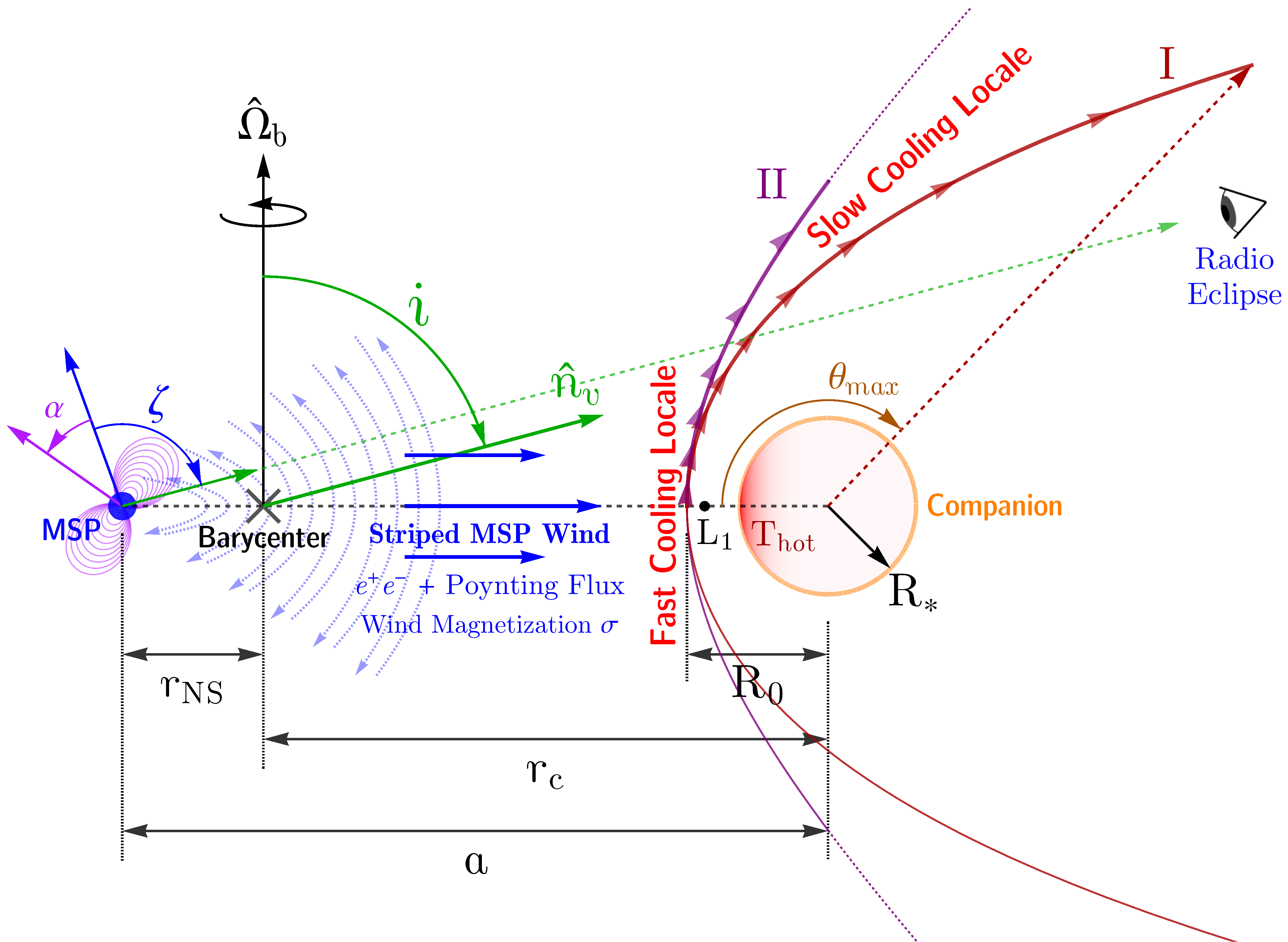}
 \vspace{-0.1truein}
\caption{Schematic cross-sectional diagram of an archetypal BW or RB, scale exaggerated for clarity, illustrating geometry of the system and observer. Note that in our model for some transitional systems and RBs that exhibit double-peaked X-ray modulation around inferior conjunction, the intrabinary shock surrounds the pulsar rather than the companion. Various geometric and model parameters that can influence the observed emission and physics are also shown that will be fully described in our upcoming paper.}
 \label{geometry_schematic}    
\end{figure}  

\vspace{-1mm}
\section{OUR MODEL}
\vspace{-1mm}

 We are developing a 3D emission model of the MSP binary systems and intrabinary shock, using the observed radio eclipses as a constraint on the spatial extent of the shock in an optically-thick formalism for radio absorption or scattering \cite{1989ApJ...342..934R}. We approximate the shocked pulsar wind and companion winds as being approximately spatially coincident, in a highly radiative limit, and azimuthally symmetric around the line joining the two masses. Analytic forms for shocks exist within this framework, for parallel-isotropic wind interaction (\cite{1996ApJ...459L..31W}, Type I) and for two colliding isotropic winds (\cite{1996ApJ...469..729C}, Type II) that are roughly borne out by MHD simulations in different contexts.  There is then a one-to-one correspondence, for a fixed radio eclipse fraction, between the shock stand-off distance $R_0$ parameter, scaling the size of the shocked region, and system inclination for a prescribed shock geometry. Orbital sweepback of the shock is treated in our upcoming paper (Wadiasingh et al. 2015 in prep), resulting in asymmetry of the eclipses and light curves, although asymmetry in radio eclipses is only pronounced at the lowest observing frequencies.  Our current focus is on the B1957+20 system, since it is the most well-studied system with deep observations spanning nearly three decades, although our model is applicable to all rotation-powered compact MSP binaries.

To guide a future particle transport analysis essential for normalizing the radiative luminosity of the shock, we compute electron cooling rates and other relevant timescales at the intrabinary shock. In Figure 2, we illustrate such timescales with the inverse Compton (IC) cooling rates computed at the stagnation point for a companion radius that is $\approx 10\%$ of the orbital separation, which is a result from optical studies of the companion in B1957+20. We assume the electron direction is tangential to the shock direction at the stagnation point, i.e. perpendicular to the line joining the pulsar and companion, for the IC cooling calculation. It is clear that synchrotron cooling dominates for most values of local magnetic field expected from the pulsar wind, for large electron Lorentz factors, confirming a similar analysis by \citep{2014A&A...561A.116B} while IC cooling can dominate at lower Lorentz factors. Even for efficient acceleration at a Larmor timescale, if a strong kilogauss companion magnetosphere is present, synchrotron burn-off may quench electron Lorentz factors to $\gamma_e \lesssim 10^7$. Moreover, for locales farther away from the stagnation point downstream in the shock, the advection timescale $\sim a/c$ dominates the transport with cooling rather slow at all but the highest Lorentz factors. 

\begin{figure}[t]
\centering
\includegraphics[scale=0.4]{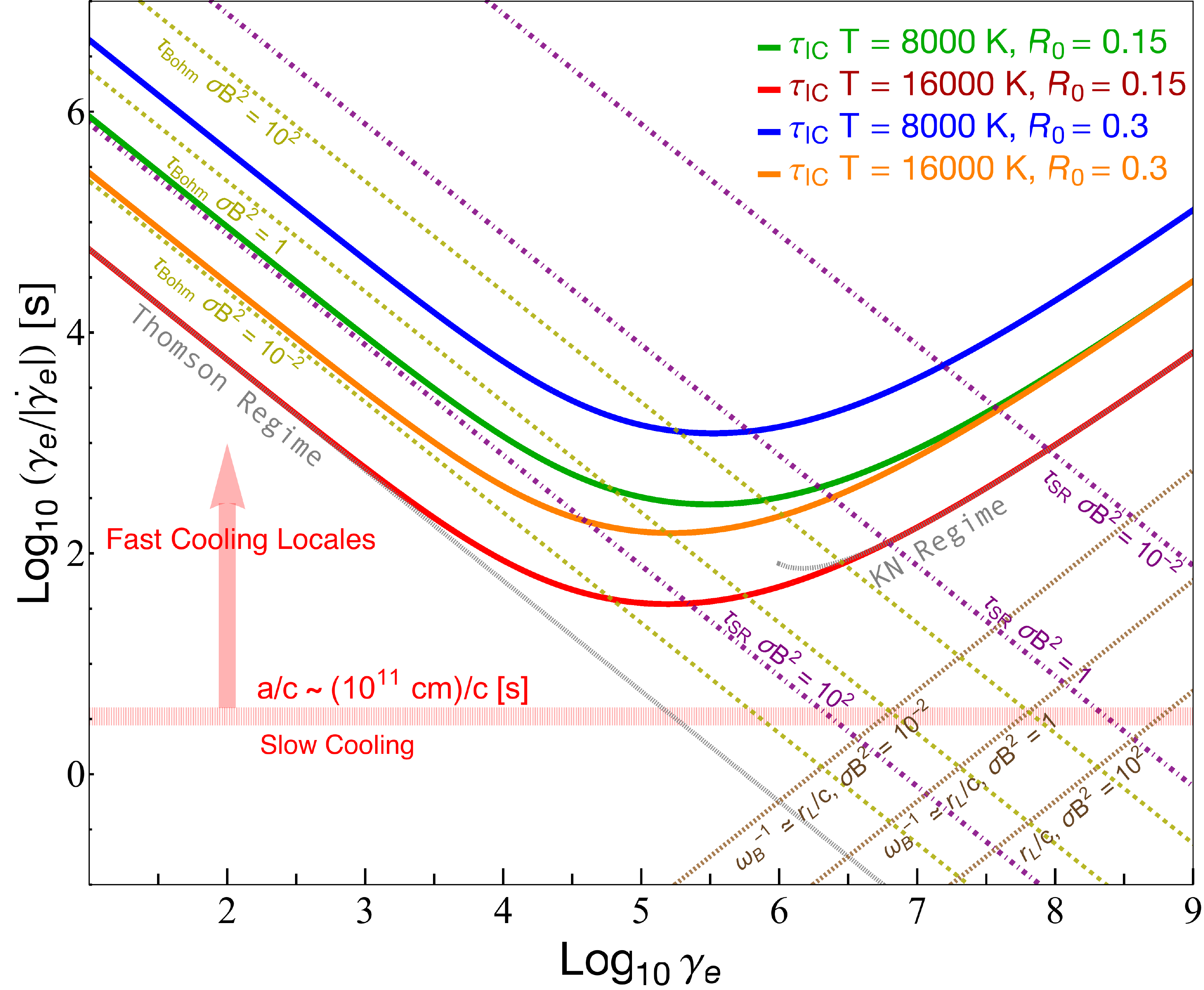}
 \vspace{-0.08truein}
\caption{Characteristic electron timescales, with cooling rates near the stagnation point $R_0$ (in units of semimajor axis length), as a function of Lorentz factor $\gamma_e$.  At high Lorentz factors, Klein-Nishina reductions to the cooling rate generate a timescale of cooling that is significantly larger than the advection timescale $\sim a/c$ or that expected from a simple Thomson extrapolation (dotted gray curve). The purple lines indicate isotropic synchrotron timescales for parameter $B_{\rm down}^2 = \sigma B^2$ the downstream local magnetic field (where $B$ is the upstream field), where $\sigma \ll1$, while the brown dotted curves demonstrate the minimum acceleration or gyro timescale $\sim r_L /c$ ($r_L$ the Larmor radius) for electrons that is smaller than $a/c$ for all but the highest Lorentz factor or lowest $\sigma B^2$ values. The dark yellow curves highlight the Bohm diffusion timescales $\tau_{\rm Bohm} \sim r_s^2 / \kappa_B$ for length scale $r_s = 2\times 10^{10}$ cm and $\kappa_{\rm B} = c r_L/3$. }
 \label{Cooling}    
\end{figure}  

Characteristic double-peaked light curves in soft X-rays are exhibited by many BWs and RBs, either centered at superior or inferior conjunction. Although geometry shadowing by a bloated companion \cite{2011ApJ...742...97B} may explain some X-ray light curves, it cannot readily explain those double-peaked light curves centered around inferior conjunction in some RBs (e.g., J2129-0429, \cite{2015arXiv150207208R}). A natural explanation for such double-peaked emission in most BWs and RBs is Doppler-boosted emission from a mildly relativistic flow, likely a shocked pulsar wind rather than the nonrelativistic companion wind, along the intrabinary shock directed towards the observer. The shock may surround the companion or pulsar, the latter case not unexpected in transitional RB systems where an accretion-powered state may have occurred in the recent past.

 For a prescribed shock geometry, stagnation point and quasi-isotropic power law distribution of electrons with index $p$, we compute light curves by calculating the Doppler factor $\delta_D$ at each point along the shock relative to the observer, and spatially integrating all contributions over the shock for each instantaneous orbital phase. The Doppler-boosting factor at each point is then $\delta_D^{2+(p-1)/2}$ for wherever the power law synchrotron emission formalism is valid, with the electrons quasi-isotropic in the comoving frame of the shock bulk flow. For Figure 3, we compute flux ratios of synchrotron emission where the flow accelerates linearly from zero velocity at the stagnation point to a maximum of $0.9$c tangent to the shock at $\theta_{\rm max} = \pi/2$; a constant magnetic field is assumed for simplicity. More sophisticated variations of parameters and assumptions are readily amenable and will be assessed in our upcoming paper. The simulated light curve morphology and peak separation depends crucially on the prescribed shock geometry and system inclination. The structure of the light curves qualitatively matches those measured for B1957+20 \cite{2012ApJ...760...92H}, with the magnitude of the flux modulation achievable by either modifying the electron density along the shock or by introducing a orbitally unmodulated DC component. The second set of curves at a lower modulation level in Figure~3, with $\varsigma(\theta)$ are generated by assuming a monotonically decreasing electron density distribution profile away from the stagnation point in the shock, the full details of which shall be discussed in the full paper. If such a DC component is thermal emission from the pulsar surface, then spectral hardening should be evident at higher energies and phases where the orbitally-modulated power law component's influence grows over the thermal component. Shrouding by the bloated companion is insufficient to explain the double-peaked structure unless the shock is extremely close to the companion surface, which is not consistent with the optically-thick formalism for radio eclipses, IR/optical companion size constraints, and the constrained inclination $i \approx 65^\circ$ of B1957+20. 
 
 \begin{figure}[t]
\twofigureoutpdf{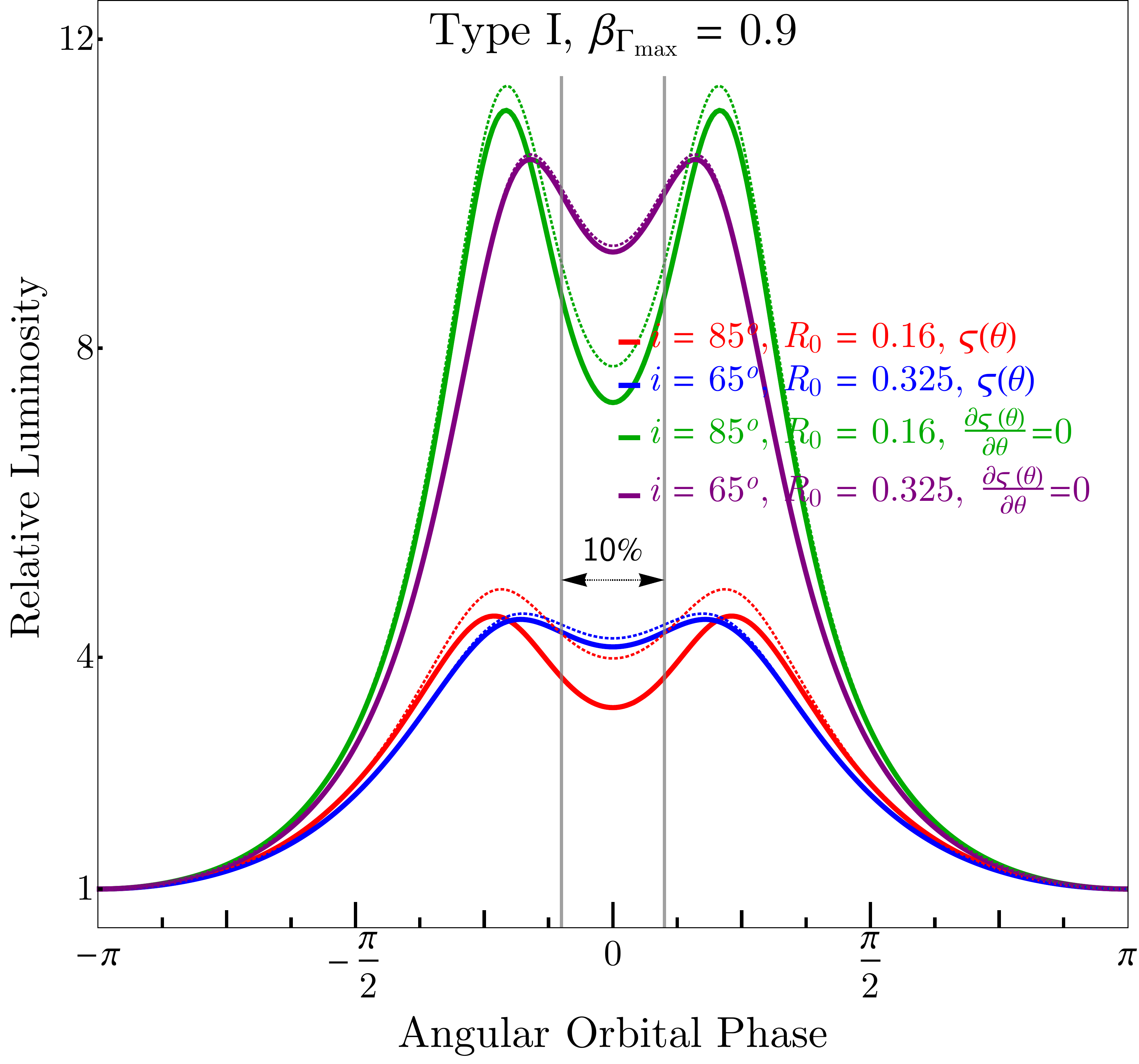}{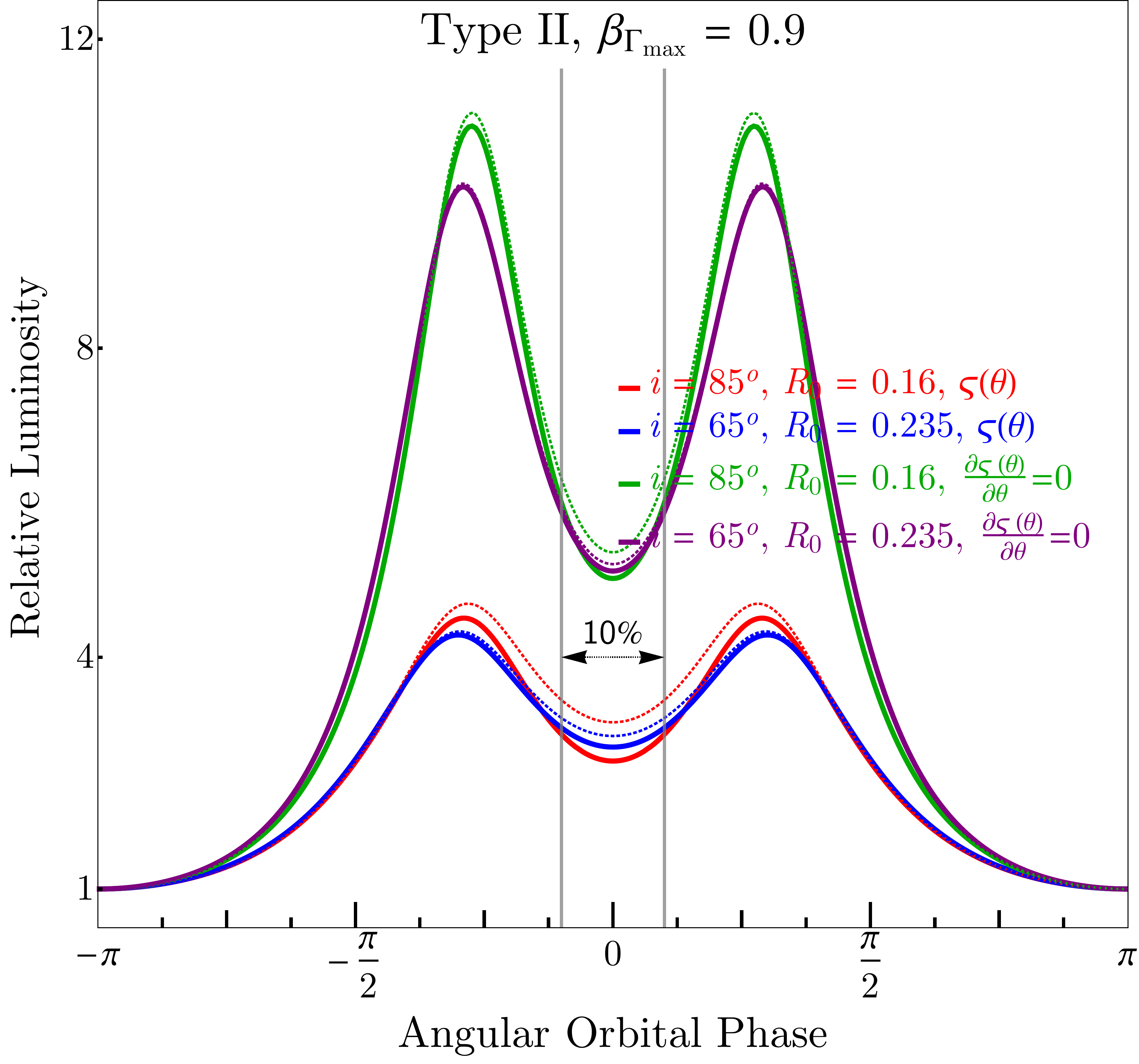}{ 
Orbitally-modulated synchrotron flux ratios of superior-to-inferior conjunction for B1957+20 with $\beta_{\Gamma \rm max} = 0.9$ the maximum speed in units of $c$ at $\theta_{\rm max}$, at an arbitrary energy where the power-law synchrotron approximation is valid (the value of B is fixed, but arbitrary here), for different inclinations and shock stand-off $R_0$ estimated from radio eclipses at superior conjunction ($10\%$ phase shown).  The solid and dashed curves are shadowed and unshadowed by the companion cases, respectively.  }  
\end{figure}  

\section{FUTURE} 

After modeling of the injection and transport of electrons along the shock to ascertain the spatial and energy evolution of the electron spectrum, we will estimate fluxes, normalize and parametrize light curves, and form model SEDs. Assessing the synchrotron self-absorption optical depth may aid in assessing any potential nonthermal optical (possibly significantly polarized) synchrotron components. We note that almost all of the double-peaked X-ray light curves of BWs and RBs are also somewhat asymmetric, with the second peak flux at a lower level than the first peak. This may be due to absorption by the ``cometary'' wind and asymmetries in the shock due to orbital sweepback, influences that we will assess and constrain. Predictions of the energy-dependent morphology of the light curve structure is also anticipated, and will serve as a useful diagnostic for system geometry and the underlying particle distribution.

 IC emission from various target photon fields present in the system may also be significant, especially from electrons present in the upstream pulsar wind that have not synchrotron cooled in the shock. For some extreme cases (e.g., J1311-3430, \cite{2015ApJ...804..115R}) where the companion flares, the IC optical depth for target optical companion photons on orbital length scales may exceed unity, yielding an observable $\gamma$-ray (orbitally modulated) signal correlated with such activity if the accelerated population of electrons contains a significant fraction of the pulsar wind energy. We encourage deeper observations and searches for orbitally-modulated emission and correlations between different wavebands for BWs and RBs.

\vspace{-0mm}
\acknowledgments
\vspace{-1mm}
The work of M.B. is supported by the South African Research Chairs Initiative of the Department of Science and Technology and the National Research Foundation of South Africa\footnote{Any opinion, finding and conclusion or recommendation expressed in this material is that of the authors and the NRF does not accept any liability in this regard.}. C.V. \& Z.W. are supported by the South African National Research Foundation. A.K.H. acknowledges support from the
NASA Astrophysics Theory Program. A.K.H., Z.W., and C.V. also acknowledge support from the \emph{Fermi} Guest Investigator Cycle 8 Grant.

\newcommand{\vol}[2]{$\,$\bf #1\rm , #2.}    
\def\mn{M.N.R.A.S.}
\def\aassupp{{Astron. Astrophys. Supp.}}
\def\apss{{Astr. Space Sci.}}
\def\apj{ApJ}
\def\nat{Nature}
\def\aaps{{Astron. \& Astr. Supp.}}
\def\aap{{A\&A}}
\def\apjs{{ApJS}}
\def\sp{{Solar Phys.}}
\def\jgr{{J. Geophys. Res.}}
\def\jphysb{{J. Phys. B}}
\def\ssr{{Space Science Rev.}}
\def\araa{{Ann. Rev. Astron. Astrophys.}}
\def\nature{{Nature}}
\def\asr{{Adv. Space. Res.}}
\def\rmp{{Rev. Mod. Phys.}}
\def\prc{{Phys. Rev. C}}
\def\prd{{Phys. Rev. D}}
\def\pr{{Phys. Rev.}}

\bibliographystyle{JHEP}
\bibliography{refsHEASA}

\providecommand{\href}[2]{#2}\begingroup\raggedright\begin{thebibliography}{10}

\bibitem{1988Natur.333..237F}
A.~S. {Fruchter}, D.~R. {Stinebring}, and J.~H. {Taylor}, {\it {A millisecond
  pulsar in an eclipsing binary}},  {\em \nat} {\bf 333} (May, 1988) 237--239.

\bibitem{1988Natur.333..832P}
E.~S. {Phinney}, C.~R. {Evans}, R.~D. {Blandford}, and S.~R. {Kulkarni}, {\it
  {Ablating dwarf model for eclipsing millisecond pulsar 1957 + 20}},  {\em
  \nat} {\bf 333} (June, 1988) 832--834.

\bibitem{2011AIPC.1357..127R}
M.~S.~E. {Roberts}, {\it {New Black Widows and Redbacks in the Galactic
  Field}},  in {\em American Institute of Physics Conference Series}
  (M.~{Burgay}, N.~{D'Amico}, P.~{Esposito}, A.~{Pellizzoni}, and
  A.~{Possenti}, eds.), vol.~1357 of {\em American Institute of Physics
  Conference Series}, pp.~127--130, Aug., 2011.
\newblock \href{http://arxiv.org/abs/1103.0819}{{\tt arXiv:1103.0819}}.

\bibitem{2009Sci...324.1411A}
A.~M. {Archibald}, I.~H. {Stairs}, S.~M. {Ransom}, V.~M. {Kaspi}, V.~I.
  {Kondratiev}, D.~R. {Lorimer}, M.~A. {McLaughlin}, J.~{Boyles}, J.~W.~T.
  {Hessels}, R.~{Lynch}, J.~{van Leeuwen}, M.~S.~E. {Roberts}, F.~{Jenet},
  D.~J. {Champion}, R.~{Rosen}, B.~N. {Barlow}, B.~H. {Dunlap}, and R.~A.
  {Remillard}, {\it {A Radio Pulsar/X-ray Binary Link}},  {\em Science} {\bf
  324} (June, 2009) 1411--1414, [\href{http://arxiv.org/abs/0905.3397}{{\tt
  arXiv:0905.3397}}].

\bibitem{2013Natur.501..517P}
A.~{Papitto}, C.~{Ferrigno}, E.~{Bozzo}, N.~{Rea}, L.~{Pavan}, L.~{Burderi},
  M.~{Burgay}, S.~{Campana}, T.~{di Salvo}, M.~{Falanga}, M.~D.
  {Filipovi{\'c}}, P.~C.~C. {Freire}, J.~W.~T. {Hessels}, A.~{Possenti}, S.~M.
  {Ransom}, A.~{Riggio}, P.~{Romano}, J.~M. {Sarkissian}, I.~H. {Stairs},
  L.~{Stella}, D.~F. {Torres}, M.~H. {Wieringa}, and G.~F. {Wong}, {\it {Swings
  between rotation and accretion power in a binary millisecond pulsar}},  {\em
  \nat} {\bf 501} (Sept., 2013) 517--520,
  [\href{http://arxiv.org/abs/1305.3884}{{\tt arXiv:1305.3884}}].

\bibitem{1990ApJ...358..561H}
A.~K. {Harding} and T.~K. {Gaisser}, {\it {Acceleration by pulsar winds in
  binary systems}},  {\em \apj} {\bf 358} (Aug., 1990) 561--574.

\bibitem{1993ApJ...403..249A}
J.~{Arons} and M.~{Tavani}, {\it {High-energy emission from the eclipsing
  millisecond pulsar PSR 1957+20}},  {\em \apj} {\bf 403} (Jan., 1993)
  249--255.

\bibitem{2012Natur.482..507A}
F.~A. {Aharonian}, S.~V. {Bogovalov}, and D.~{Khangulyan}, {\it {Abrupt
  acceleration of a `cold' ultrarelativistic wind from the Crab pulsar}},  {\em
  \nat} {\bf 482} (Feb., 2012) 507--509.

\bibitem{2007MNRAS.379.1117R}
M.~T. {Reynolds}, P.~J. {Callanan}, A.~S. {Fruchter}, M.~A.~P. {Torres}, M.~E.
  {Beer}, and R.~A. {Gibbons}, {\it {The light curve of the companion to PSR
  B1957+20}},  {\em \mnras} {\bf 379} (Aug., 2007) 1117--1122,
  [\href{http://arxiv.org/abs/0705.2514}{{\tt arXiv:0705.2514}}].

\bibitem{2013ApJ...769..108B}
R.~P. {Breton}, M.~H. {van Kerkwijk}, M.~S.~E. {Roberts}, J.~W.~T. {Hessels},
  F.~{Camilo}, M.~A. {McLaughlin}, S.~M. {Ransom}, P.~S. {Ray}, and I.~H.
  {Stairs}, {\it {Discovery of the Optical Counterparts to Four Energetic Fermi
  Millisecond Pulsars}},  {\em \apj} {\bf 769} (June, 2013) 108,
  [\href{http://arxiv.org/abs/1302.1790}{{\tt arXiv:1302.1790}}].

\bibitem{2014ApJS..213....6J}
T.~J. {Johnson}, C.~{Venter}, A.~K. {Harding}, L.~{Guillemot}, D.~A. {Smith},
  M.~{Kramer}, {\"O}.~{{\c C}elik}, P.~R. {den Hartog}, E.~C. {Ferrara},
  X.~{Hou}, J.~{Lande}, and P.~S. {Ray}, {\it {Constraints on the Emission
  Geometries and Spin Evolution of Gamma-Ray Millisecond Pulsars}},  {\em
  \apjs} {\bf 213} (July, 2014) 1--54,
  [\href{http://arxiv.org/abs/1404.2264}{{\tt arXiv:1404.2264}}].

\bibitem{1989ApJ...342..934R}
F.~A. {Rasio}, S.~L. {Shapiro}, and S.~A. {Teukolsky}, {\it {What is causing
  the eclipse in the millisecond binary pulsar?}},  {\em \apj} {\bf 342} (July,
  1989) 934--939.

\bibitem{1996ApJ...459L..31W}
F.~P. {Wilkin}, {\it {Exact Analytic Solutions for Stellar Wind Bow Shocks}},
  {\em \apjl} {\bf 459} (Mar., 1996) L31.

\bibitem{1996ApJ...469..729C}
J.~{Canto}, A.~C. {Raga}, and F.~P. {Wilkin}, {\it {Exact, Algebraic Solutions
  of the Thin-Shell Two-Wind Interaction Problem}},  {\em \apj} {\bf 469}
  (Oct., 1996) 729.

\bibitem{2014A&A...561A.116B}
W.~{Bednarek}, {\it {Modulated gamma-ray emission from compact millisecond
  pulsar binary systems}},  {\em \aap} {\bf 561} (Jan., 2014) A116, 1--9,
  [\href{http://arxiv.org/abs/1311.7505}{{\tt arXiv:1311.7505}}].

\bibitem{2011ApJ...742...97B}
S.~{Bogdanov}, A.~M. {Archibald}, J.~W.~T. {Hessels}, V.~M. {Kaspi},
  D.~{Lorimer}, M.~A. {McLaughlin}, S.~M. {Ransom}, and I.~H. {Stairs}, {\it {A
  Chandra X-Ray Observation of the Binary Millisecond Pulsar PSR J1023+0038}},
  {\em \apj} {\bf 742} (Dec., 2011) 97,
  [\href{http://arxiv.org/abs/1108.5753}{{\tt arXiv:1108.5753}}].

\bibitem{2015arXiv150207208R}
M.~S.~E. {Roberts}, M.~A. {McLaughlin}, P.~A. {Gentile}, P.~S. {Ray}, S.~M.
  {Ransom}, and J.~W.~T. {Hessels}, {\it {X-Ray Studies of Redbacks}},  {\em
  ArXiv e-prints} (Feb., 2015) [\href{http://arxiv.org/abs/1502.07208}{{\tt
  arXiv:1502.07208}}].

\bibitem{2012ApJ...760...92H}
R.~H.~H. {Huang}, A.~K.~H. {Kong}, J.~{Takata}, C.~Y. {Hui}, L.~C.~C. {Lin},
  and K.~S. {Cheng}, {\it {X-Ray Studies of the Black Widow Pulsar PSR
  B1957+20}},  {\em \apj} {\bf 760} (Nov., 2012) 92,
  [\href{http://arxiv.org/abs/1209.5871}{{\tt arXiv:1209.5871}}].

\bibitem{2015ApJ...804..115R}
R.~W. {Romani}, A.~V. {Filippenko}, and S.~B. {Cenko}, {\it {A Spectroscopic
  Study of the Extreme Black Widow PSR J1311-3430}},  {\em \apj} {\bf 804}
  (May, 2015) 115, 1--10, [\href{http://arxiv.org/abs/1503.05247}{{\tt
  arXiv:1503.05247}}].

\end{thebibliography}\endgroup

\end{document}